# Stochastic dynamics of macromolecular-assembly networks

Leonor Saiz and Jose M. G. Vilar[*]

Integrative Biological Modeling Laboratory, Computational Biology Program, Memorial Sloan-Kettering Cancer Center, 1275 York Avenue, Box #460, New York, NY 10021, USA

[*]Correspondence: vilar@cbio.mskcc.org

**Running title:** Stochastic macromolecular assembly

**Subject Categories:** Computational methods, Metabolic and regulatory networks

**Corresponding author:**

Jose M.G. Vilar

Computational Biology Center

Memorial Sloan-Kettering Cancer Center

307 East 63rd Street

New York, NY 10021, USA

Tel: 646-735-8085

Fax: 646-735-0021

E-mail: vilar@cbio.mskcc.org

**Manuscript information:** 40 text pages and 5 Figures.

Total number of characters: 50,112 (59,579 with spaces).

## Abstract

The formation and regulation of macromolecular complexes provides the backbone of most cellular processes, including gene regulation and signal transduction. The inherent complexity of assembling macromolecular structures makes current computational methods strongly limited for understanding how the physical interactions between cellular components give rise to systemic properties of cells. Here we present a stochastic approach to study the dynamics of networks formed by macromolecular complexes in terms of the molecular interactions of their components. Exploiting key thermodynamic concepts, this approach makes it possible to both estimate reaction rates and incorporate the resulting assembly dynamics into the stochastic kinetics of cellular networks. As prototype systems, we consider the *lac* operon and phage $\lambda$ induction switches, which rely on the formation of DNA loops by proteins and on the integration of these protein-DNA complexes into intracellular networks. This cross-scale approach offers an effective starting point to move forward from network diagrams, such as those of protein-protein and DNA-protein interaction networks, to the actual dynamics of cellular processes.

## Introduction

Cells consist of thousands of different molecular species that orchestrate their interactions to form extremely reliable functional units (Hartwell et al., 1999). Such molecular diversity and the pervasive ability of cellular components to establish multiple simultaneous interactions typically lead to the formation of large heterogeneous macromolecular assemblies, also known as complexes (Pawson and Nash, 2003). These complexes form the backbone of the most fundamental cellular processes, including gene regulation and signal transduction. Important examples are the assembly of the eukaryotic transcriptional machinery (Roeder, 1998), with about hundred components, and the formation of signaling complexes (Bray, 1998), with tens of different molecular species.

One of the main challenges facing modern biology is to move forward from the reductionist approach into the systemic properties of biological systems (Alon, 2003). A major goal is to understand how the dynamics of cellular processes emerges from the interactions among the different molecular components. Typical computational approaches approximate cellular processes by networks of chemical reactions between different molecular species (Endy and Brent, 2001). A strong barrier to this type of approaches is the inherent complexity of macromolecular complex formation. Complexes are typically made of smaller building blocks with a modular organization that can be combined in a number of different ways (Pawson and Nash, 2003). The result of each combination is a specific molecular species and should be considered explicitly in a chemical reaction description. Therefore, there are potentially as many reactions as the number of possible ways of arranging the different elements, which grows exponentially with the number of the constituent elements. Twenty components, for instance, already give rise to over a million of possible species.

Two main types of avenues have been followed to tackle the exponential growth of the number of molecular species that arise during macromolecular assembly. One is based on computer programs that generate reaction rate equations for all of the macromolecular species (Bray and Lay, 1997). The other generates the different species dynamically (Lok and Brent, 2005; Morton-Firth and Bray, 1998). Yet, none of the existing methods provides a consistent way to estimate the reaction rates. This obstacle is remarkable because the potential number of rates is even higher than the number of possible complexes. As a result, those methods often lead to unrealistic situations, such as the formation of polymeric complexes that do not exist under physiological conditions, which has been noted explicitly as intriguing caveats of the existing methodologies (Bray and Lay, 1997; Lok and Brent, 2005).

Here we develop a stochastic approach based on the underlying physico-chemistry of macromolecular assembly that brings the properties of macromolecular interactions across scales up to the dynamics of the cellular networks. This approach not only generates the molecular species dynamically but, at the same time, also estimates the needed reaction rates in a way that is fully consistent with the underlying thermodynamics. In this way, it does not give rise to the formation of unrealistic complexes and it addresses the exponential growth of the number of species by stochastically exploring the set of representative complexes. To illustrate the applicability of the approach, we focus on DNA-protein complexes and their integration in gene regulatory networks. We consider as prototype systems the induction switches in the *lac* operon (Müller-Hill, 1996) and phage λ (Ptashne, 2004), the two systems that led to the discovery of gene regulation (Jacob and Monod, 1961).

## Representation of macromolecular complexes

A crucial aspect is to use a description for macromolecular complexes that can capture the underlying complexity in simple terms. It is possible to take advantage of the fact that macromolecular complexes have typically unambiguous structures, where only certain molecular species can occupy a given position within the complex. In such cases, the specific configuration or state of the macromolecular complex can be described by a set of $M$ variables, denoted by $s = (s_1, \ldots s_i, \ldots s_M)$, whose values indicate whether a particular molecular component is present or not at a specific position. We chose $s_i = 1$ to indicate that the component is present and $s_i = 0$ to indicate that it is not present (Figure 1a). With this description, the potential number of specific complexes is $2^M$ and the number of reactions, $\frac{1}{2} 2^M (2^M - 1) \approx 2^{2M-1}$.

The use of binary variables provides a concise method to describe all the potential complexes without explicitly enumerating them. This type of approach has been used in a wide range of interesting biological situations, such as diverse allosteric processes (Bray and Duke, 2004), binding of molecules to a substrate (Di Cera, 1995; Keating and Di Cera, 1993), binding of multi-state proteins to receptor docking sites (Borisov et al., 2005), and signaling through clusters of receptors (Bray et al., 1998; Duke and Bray, 1999). In practice, current binary-variable approaches have strong limitations to tackle the assembly of macromolecular complexes. On the one hand, there are combinations of variables that do not have a physical existence. Explicitly, if a component that bridges two disconnected parts of the complex is missing, then the complex does not exist and if two components occupy the same position, they cannot be present within the complex simultaneously. On the other hand, the structure of the complex does not have to remain fixed. A complex can have different conformations and the components can be present in several states with different properties. None of the existing approaches based on binary variables incorporates all these key features needed to study macromolecular complex formation. In the following section, we exploit the underlying thermodynamics to put forward a binary description applicable to macromolecular assembly.

## Thermodynamics of assembly

Thermodynamics allows for a straightforward connection of the binary description with the molecular properties of the system. Each configuration of the macromolecular complex has a corresponding free energy, which is a quantity that indicates the tendency of the system to change its state. Transformations that decrease the free energy of the system are favored over those that increase it; i.e., the tendency of the system is to evolve towards the lowest free energy minima.

The statistical interpretation of thermodynamics (Gibbs, 1902; Hill, 1960) connects the equilibrium probability $P_s$ of state $s$ with its free energy $\Delta G(s)$ through the expression

$$P_s = \frac{1}{Z} e^{-\Delta G(s)/RT},$$

where $Z = \sum_s e^{-\Delta G(s)/RT}$ is the normalization factor and $RT$ is the gas constant times the absolute temperature ($RT \approx 0.6$ kcal/mol for typical experimental conditions).

One crucial aspect is that the free energy of any given configuration, $s = (s_1,...s_i,...s_N)$, can be obtained to any degree of accuracy by expanding the free energy in powers of the binary variables:

$$\Delta G(s) = \Delta G(0) + \sum_{i=1}^{M} \Delta G_i s_i + \sum_{i=1}^{M-1} \sum_{j=i+1}^{M} \Delta G_{ij} s_i s_j + \sum_{i=1}^{M-2} \sum_{j=i+1}^{M-1} \sum_{k=j+1}^{M} \Delta G_{ijk} s_i s_j s_k + ...$$

The first term in the right hand side of the equation is the free energy of the empty complex, when none of their components is present, and it serves as a reference free energy; the second term takes into account the energetic cost of placing each element in the complex; the third term accounts for the pairwise interactions between elements; the fourth term accounts for interactions that need a third element to take place; and higher order terms, not shown in the equation, account for interactions that take place only when multiple elements are present. This expansion can be stopped at any degree of complexity. Typical network diagrams, such as those of protein-protein interactions, represent only up to pairwise interactions (the first three terms). The main advantage of this expansion is that the $2^M$ free energies needed to characterize all the possible configurations can be obtained from a much smaller subset.

In order to connect the lower terms of this expansion directly with experimental data, the most important piece of information is that the free energy of binding, $\Delta G_{bind}$, between two elements can be decomposed into two main contributions (Vilar and Saiz, 2005):

$$\Delta G_{bind} = \Delta G_{pos} + \Delta G_{int}\,.$$

One of them, the interaction free energy $\Delta G_{int}$, arises from the interactions between the two molecules, such as electrostatic, hydrophobic, and Van der Waals interactions (Honig and Nicholls, 1995). The other, the positional free energy $\Delta G_{pos}$, results from positioning the molecules in the right place and orientation so that they can interact and it accounts, among other potential contributions, for the loss of translational and rotational entropy upon binding (Page and Jencks, 1971).

The expression for the positional free energy can be written as

$$\Delta G_{pos} = \Delta G_{pos}^{o} - RT\ln[N]\,,$$

where the quantity $\Delta G_{pos}^{o}$ is the molar positional free energy and $[N]$ is the concentration expressed in moles ($\Delta G_{pos} = \Delta G_{pos}^{o}$, for $[N] = 1M$). In general, the free energy of binding depends on the concentrations of the different components through the positional free energy, $\Delta G_{pos}$.

Typical values of the molar positional free energy are $\Delta G_{pos}^{o} \approx 15$ kcal/mol (Finkelstein and Janin, 1989). This value has been computed theoretically and measured experimentally for simple molecules (Finkelstein and Janin, 1989; Page and Jencks, 1971). For complex proteins, it might differ slightly. For instance, for the binding of AChE-Fas2 it has been estimated to be 9 kcal/mol (Minh et al., 2005). Such high values of the positional free energy indicate that if the free energy of interaction is zero, the state in which two molecules are as close as if they were bound is extremely unlikely. Even small values of binding free energies, such as $\Delta G_{bind}^{o} \approx -2\,\text{kcal/mol}$, would imply considerably high interaction free energies, such as $\Delta G_{int}^{o} \approx -17\,\text{kcal/mol}$ (assuming $\Delta G_{pos}^{o} \approx 15\,\text{kcal/mol}$). This is a crucial point in order to properly account for disconnected complexes. If the bridging element is missing, there is only one missing positional free energy but two interaction free energies (Figure 1b and 1c). Thus the free energy of the disconnected complex (Fig. 1b) is much higher than that of the connected one (Fig. 1c), which indicates that the disconnected complex is extremely less abundant than the connected one under physiological conditions. The concept of positional free energy is also essential to avoid the presence of spurious polymeric complexes. A compact complex (Fig. 1d) will have additional free energies of interaction compared to the chain-like one (Fig. 1c), which will render it much more stable than the polymeric chain. Finally, when two elements cannot be present in the same position at the same time, with this approach their interaction energy is infinitely large and the complex does not exist in practice.

The term $\sum_{i=1}^{M-1}\sum_{j=i+1}^{M}\Delta G_{ij}s_{i}s_{j}$ accounts, among others, for interactions between components of the complex that have a multidomain structure, where domains interact in a pairwise fashion with each other. In general, as we show in detail in the examples below, one should also consider the conformational free energy that accounts for the structural changes needed to accommodate multiple simultaneous interactions. This type of interactions requires higher order terms in the free energy expansion, such as

$$\sum_{i=1}^{M-2} \sum_{j=i+1}^{M-1} \sum_{k=j+1}^{M} \Delta G_{ijk} s_i s_j s_k \; .$$

All these thermodynamic concepts blend naturally into a binary-variable description to provide an approach that incorporates the key elements needed for studying macromolecular complex assembly, such as avoiding the formation of unrealistic polymeric complexes, taking into account disconnected complexes, considering conformational changes, and incorporating mediated or multi-component interactions.

## Applications: making networks out of complexes and vice versa

The formation of DNA loops by proteins and protein complexes in the regulation of the *lac* operon and phage λ provides challenging examples to illustrate the applicability of this approach. Full understating of these two genetic systems requires the use of thermodynamic concepts not considered by current methods to study macromolecular assembly. These concepts are essential to tackle more complex situations, such as gene regulation in eukaryotes, which relies widely on DNA looping to implement action at a distance from regulatory elements that are far away from the promoter region (Roeder, 2003; Yasmin et al., 2004). In particular, decomposition of the free energy into interaction, positional, and conformational contributions is crucial for understanding how weak distal DNA binding sites can strongly affect transcription (Vilar and Leibler, 2003) and how proteins that do not oligomerize in solution do so on DNA to form DNA loops (Dodd et al., 2001; Revet et al., 1999; Vilar and Saiz, 2005). Hereafter, to simplify the notation, we will refer to positional, interaction, and conformational free energies by $p$, $e$, and $c$, respectively.

One remarkable aspect of our approach is that the binary-variable method we propose has an equivalent representation in terms of network diagrams, in which nodes indicate whether or not a component or molecular conformation is present and links represent the interactions among components and molecular conformations. These networks can be directly mapped to the underlying macromolecular structure of the complexes and their properties, as illustrated in detail below for the cases of DNA looping in the *lac* operon and phage $\lambda$.

The resulting networks strikingly resemble the underlying molecular structures because nodes are associated with properties of the components, which have a determined arrangement in space. In this type of interaction networks, the whole network specifies a single state of the system, in the same way as a state of the macromolecular assembly is specified by indicating where each component is within the complex. This component-oriented description allows for a representation that is not affected by the exponential explosion in the number of states as the number of components increases. In other widely used methods with network representations, like Markov Chains, each node represents a state of the system and therefore there are as many nodes in the network as potential states. In those state-oriented networks, only one node is occupied at a time and the behavior of the system is represented by a series of jumps from one node to another.

In our case, several nodes can be occupied simultaneously and the behavior of the system is given by the sequence of changes in occupancy. In Figure 2, we compare a general interaction network representation for a three-component complex with the graphical representation of a Markov Chain for the same system. The number of nodes in the interaction network is the same as the number of components. In the Markov Chain network, in contrast, the number of nodes is 2 to the number of components.

This avenue for incorporating binary variables with thermodynamics presents important differences with the way in which thermodynamic concepts have been used so far in macromolecular assembly. Thermodynamic concepts applied to gene regulation were pioneered by Shea and Ackers  (Ackers et al., 1982; Shea and Ackers, 1985) and subsequently used in a variety of situations (Bintu et al., 2005).  In the traditional framework, the probability for the system to be in a state $k$ is given by $P_k = \dfrac{1}{Z}[N]^{jk} e^{-\Delta G_k^o / RT}$, where  $\Delta G_k^o$ is the standard (Molar) free energy of the state $k$, $jk$ is the number of molecules bound in the state $k$, and  $Z = \sum_k [N]^{jk} e^{-\Delta G_k^o / RT}$  is the normalization factor (Hill, 1960). The summation is taken over all the states. Thus, to describe the system with the traditional approach one has to give the standard free energy for each state as well as the number of molecules bound. Typically, this is done in the form of a table, which has as many entries as the number of states of the system. Thus, for a system with 3 components there is a table with 8 entries for the standard free energy and another 8  for the number of molecules.

In the approach we have proposed, there is a simple formula that accounts for the free energy of all the states, and the resulting expressions have a compact form amenable to computational and mathematical manipulations. As we show in the following sections, this approach brings forward explicitly the connection between macromolecular structure and function and its integration into the biochemical dynamics of cellular networks.

## The *lac* operon

The *lac* operon consists of a regulatory domain and three genes required for the uptake and catabolism of lactose (Müller-Hill, 1996). Its main regulator is the *lac* repressor, which can bind to DNA at the main operator site $O1$, thus preventing transcription, and to an auxiliary operator ($O2$ or $O3$) by looping the intervening DNA.

For a system with two operators ($O1$ and $O2$), the *lac* repressor-DNA complex (Lewis et al., 1996) can be in five representative states (Vilar and Leibler, 2003): (i) none of the operators is occupied, (ii) a repressor is bound to just the auxiliary operator, (iii) a repressor is bound to just the main operator, (iv) a repressor is bound to both the main and the auxiliary operators by looping the intervening DNA, and (v) one repressor is bound to the main operator and another repressor, to the auxiliary operator.

We can express the free energy of all these states in a compact form in terms of binary variables:

$$\Delta G(s) = \Delta G(0) + p(s_1 + s_2) + p_{DNA}s_{DNA} + (e_1 s_1 + e_2 s_2)s_{DNA} + (c_L - ps_1 s_2)s_L s_{DNA} \, ,$$

where $s_1$ and $s_2$ are the binary variables that indicate whether ($s_i = 1$; for $i = 1, 2$) or not ($s_i = 0$; for $i = 1, 2$) the repressor is bound to *O1* and *O2*, respectively; $s_{DNA}$ indicates the presence ($s_{DNA} = 1$) or absence ($s_{DNA} = 0$) of DNA; and $s_L$ is a variable that indicates the molecular conformation of DNA, either looped ($s_L = 1$) or unlooped ($s_L = 0$). The quantities $p$ and $p_{DNA}$ are the positional free energies of the repressor and DNA, respectively; $e_1$ and $e_2$, the interaction free energy between the repressor and *O1* and *O2*, respectively; and $c_L$, the conformational free energy of looping DNA. $\Delta G(0)$ is the free energy of the reference state in which there is no repressor, DNA, or DNA looping (all the binary variables are zero).

We are interested in the binding of the repressor to DNA and therefore DNA is always present ($s_{DNA} = 1$), which simplifies the description to just three binary variables:

$$\Delta G(s) = (p + e_1)s_1 + (p + e_2)s_2 + (c_L - ps_1 s_2)s_L,$$

where we have chosen $\Delta G(0) = -p_{DNA}$ so that the reference free energy is equal to zero when no repressor is bound and there is no DNA looping. The network representation of the *lac* repressor-DNA assembly with the three binary variables (Fig. 3a) has close connection with the underlying molecular structural properties (Fig. 3b), in such a way that, given the structural arrangement of a complex, our approach provides a straightforward avenue to obtain a network representation with an associated thermodynamic description. Explicitly, each node in the interaction network corresponds to a binary variable in the equation for the free energy.

The decomposition of the free energy in its different contributions (positional, interaction, and conformational) becomes crucial to perform an expansion in terms of binary variables. From the statistical thermodynamics point of view, the binding of the repressor to two operators has only five relevant states (*i–v*). With the binary description of the state of the complex there are three variables and therefore $2^3 = 8$ states. These two descriptions are in fact equivalent because three of the eight states in the binary description have considerably high free energies, i.e., extremely low probabilities, which for practical purposes makes them irrelevant (see Fig. 3c). The high free energies arise because for these states positional and conformational free energies are not balanced by interaction free energies.

This approach can naturally be used to study the consequences that DNA looping has in gene regulation. In the *lac* operon, transcription takes place only when the main operator *O1* is free; i.e., when the binary variable $s_1$ is zero. Thus, the transcription rate, $\tau$, is proportional to the average value of 1 minus $s_1$:

$$\tau = \frac{1}{Z} \sum_s \tau_{\max} (1 - s_1) e^{-\Delta G(s)/RT} \, ,$$

where $\tau_{\max}$ is the maximum transcription rate. This model shows a precise agreement with experiments (Oehler et al., 1994) over the three orders of magnitude of the measured repression levels (Fig. 3d). The specific form that DNA looping confers to the repression level as a function of the repressor concentration has two notable characteristics. The repression level has both a significantly high value and a relatively flat profile around physiological *lac* repressor concentrations (~15 nM). Both properties have important general consequences for the underlying microbiochemistry of the cell. If concentrations of the different molecular species are kept low to prevent non-specific interactions, not only is the binding to the specific sites decreased but also fluctuations are expected to become important (Elowitz et al., 2002; Paulsson, 2004; Rosenfeld et al., 2005). In the case of the *lac* operon, the average number of repressors per cell is very low, around 10, and, because of this low value, is expected to fluctuate strongly from cell to cell. The effects of DNA looping, as exemplified by the repression level, not only increase specificity and affinity of the *lac* repressor for the main operator but, at the same time, also make transcription fairly insensitive to fluctuations in the number of repressors.

We can address a more general situation, which includes the binding of different molecular species at identical positions within the complex. To illustrate this possibility, we consider the case of mutant *lac* repressors that do not tetramerize (Oehler et al., 1990). In its dimeric form the *lac* repressor has just a single DNA binding domain and thus cannot loop DNA. In this case, the free energy is given by $\Delta G(s) = (p_d + e_1)s_{1d} + (p_d + e_2)s_{2d}$, where $p_d$ is the positional free energy of the dimeric *lac* repressor and $s_{1d}$ and $s_{2d}$ are the two binary variables that account for its binding to *O1* and *O2*, respectively. If this mutant repressor is expressed together with the wild type (WT) *lac* repressor, it will compete for the binding to the operator sites and the free energy will be given by the sum of the free energies for WT and mutant repressors plus a contribution accounting for the interaction between the two types of repressors:

$$\Delta G(s) = (p + e_1)s_1 + (p + e_2)s_2 + (c_L - ps_1s_2)s_L +$$
$$(p_d + e_1)s_{1d} + (p_d + e_2)s_{2d} + \infty(s_1s_{1d} + s_2s_{2d}).$$

The term $\infty(s_1s_{1d} + s_2s_{2d})$ introduces an infinite free energy of interaction when the two types of repressors are bound to the same operator simultaneously, thus making the probability of such states zero. The five-binary-variable description of the WT and mutant *lac* repressor-DNA complex formation is an straightforward extension of the one for just WT *lac* repressor. This example clearly illustrates how it is possible to use our approach to add complexity without escalating into an exponentially growing description.

## Phage λ

The genetic regulation of phage λ provides an explicit example in which the DNA loop is formed not by a single protein, as in the *lac* operon, but by a protein complex that is assembled on DNA as the loop forms. The lysogenic to lytic switch in phage λ infected *Escherichia coli* lysogens is controlled at two operators in the phage DNA. These two operators, known as the left, $O_L$, and right, $O_R$, operators, are located 2.4 kb away from one another. Each of them has a tandem of three DNA motifs where phage λ cI repressors can bind as dimers (cI$_2$): r1, r2, and r3 for the right operator; and l1, l2, and l3 for the left operator. Two cI dimers bound to r1 and r2 on the right operator can form an octamer with two cI dimers bound to l1 and l2 on the left operator by looping the intervening DNA.

Stability of the *E. coli* lysogens is accomplished by repression of transcription by the phage λ cI repressor of the *cro* gene at the P$_r$ promoter and regulation of its own transcription at the P$_{rm}$ promoter. Explicitly, binding of cI repressor dimers to r2, when r3 is vacant, activates its own transcription. When r3 is occupied, *cI* transcription is turned off. For a long time, one of the main puzzles in the regulation of phage λ was that the strength of r3 was too weak for it to be occupied by cI at the observed physiological concentrations (Ptashne, 2004). The missing element was that the formation of the DNA loop by the octamerization of the cI repressor dimers bound at r1, r2, l1, l2 can bring l3 close to r3 so that the cI repressor can bind cooperatively as a tetramer to these two sites even though they are ~2.4 kb apart.

Modeling of this system already gets close to the limits of the traditional thermodynamic approach. The number of states is 128, which accounts for all the combinations of occupancies of the six binding sites in either the looped or unlooped conformations of DNA. In terms of binary variables, however, the free energy of all the possible states of the assembly of the cI repressor-DNA complex is simply described by

$$\Delta G(s) = \sum_{i=1}^{3} (p + e_{ri})s_{ri} + e_{r23}s_{r2}s_{r3} + e_{r12}s_{r1}s_{r2} + e_{r123}s_{r1}s_{r2}s_{r3}$$
$$+ \sum_{i=1}^{3} (p + e_{li})s_{li} + e_{l23}s_{l2}s_{l3} + e_{l12}s_{l1}s_{l2} + e_{l123}s_{l1}s_{l2}s_{l3}$$
$$+ \left( c_L + e_T s_{r3}s_{l3} + e_O s_{r2}s_{l2}s_{r1}s_{l1} \right)s_L .$$

Here, $p$ is the positional free energy of the cI repressor dimers; and $e_{ri}$ and $e_{li}$, with $i = 1, 2, 3$, are the interaction free energy of the cI dimer with each of its three DNA binding sites at the right and left operators, respectively. The terms of the type $e_{r23}s_{r2}s_{r3}$ account for the pairwise interactions of cI dimers bound at neighboring DNA sites. The terms with three binary variables, such as $e_{r123}s_{r1}s_{r2}s_{r3}$, are introduced because these pairwise interactions are affected by the binding of cI repressors to the other neighboring site. The term $\left(c_L + e_T s_{r3} s_{l3} + e_O s_{r2} s_{l2} s_{r1} s_{l1}\right) s_L$ accounts for the effects of DNA looping. It includes looping ($c_L$), cI tetramerization ($e_T$), and cI octamerization ($e_O$) contributions to the free energy. This system has a network representation (Fig. 4a) with seven binary variables that closely follows from its molecular organization (Fig. 4b). As shown here, the seven binary-variable description is equivalent to previous modeling of this system using the traditional thermodynamic approach (Dodd et al., 2004).

It is important to note that the expression, $e_{r23} s_{r2} s_{r3} + e_{r12} s_{r1} s_{r2} + e_{r123} s_{r1} s_{r2} s_{r3}$, can be rewritten in the more familiar form $e_{r23} s_{r2} s_{r3}(1 - s_{r1}) + e_{r12} s_{r1} s_{r2}(1 - s_{r3}) + (e_{r123} + e_{r12} + e_{r23}) s_{r1} s_{r2} s_{r3}$, which indicates that the free energy of the cooperative interactions between all the cI dimers bound to the right operator is $e_{r123} + e_{r12} + e_{r23}$. This expression explicitly reveals $e_{r123}$ as the correction to the free energy from the effects of a third cI dimer on the intra-operator interactions between pairs of cI dimers and illustrates how it is possible to expand the free energy of a given configuration, $s = (s_1, \ldots s_i, \ldots s_M)$, in powers of the binary variables.

In general, there is the potential for establishing multiple loops. In the traditional approach, considering two additional loops will increase the number of states from 128 to 256. In our case, this extension involves adding just three terms, $\left(c_{L2} + e_O s_{r2} s_{l3} s_{r1} s_{l2}\right) s_{L2} + \left(c_{L3} + e_O s_{r3} s_{l2} s_{r2} s_{l1}\right) s_{L3} + \infty(s_L s_{L2} + s_L s_{L3} + s_{L2} s_{L3})$, to the free energy $\Delta G(s)$. Here, $s_{L2}$ and $s_{L3}$ are the binary variables for the two additional loops and $c_{L2}$ and $c_{L3}$ are the corresponding free energies of looping. For the wild type situation, the probability of having these extra loops is very small and they do not significantly affect the behavior of the system (Dodd et al., 2004). However, they might become important in mutants with altered binding.

The effects of the left operator and DNA looping in the induction switch of phage λ are apparent in the transcription rate at the $P_{rm}$ promoter, which depends strongly on the occupancy of r2 and r3. There is transcription at a basal level $\tau_{bas}$ when neither r2 nor r3 are occupied and at an activated level $\tau_{act}$ when r2 is occupied and r3 is free. The activated transcription rate, in turn, depends on whether ($\tau_{act} = \tau_{actl}$) or not ($\tau_{act} = \tau_{actnl}$) DNA is looped (Hochschild and Ptashne, 1988). These dependences are expressed mathematically through

$$\tau = \frac{1}{Z}\sum_s ((\tau_{actnl}(1-s_L) + \tau_{actl}s_L)s_{r2} + \tau_{bas})(1-s_{r3})e^{-\Delta G(s)/RT} .$$

The activity of $P_{rm}$ as a function of the cI repressor concentration shows a sharp maximum for wild type phage DNA and a plateau-like maximum for two mutants in which the r3 site is not occupied at WT concentrations (Fig. 4c). One of these mutants has only the right operator and the other has a weak l3 site. The narrower maximum of wild type allows for tighter control of the cI concentration, whose production sharply decreases for high concentrations. This marked decrease is the result of the extra layer of cooperativity of binding to r3 introduced by DNA looping and has important consequences for the kinetics of the system.

## Stochastic dynamics and macromolecular assembly networks

A relationship between the kinetics and the thermodynamic properties of the system can be exploited to infer transition rates. It is known as the principle of detailed balance and results from the fact that at equilibrium the rates of going from state *s* to state *s'* and its inverse, from *s'* to *s*, are the same. Mathematically, it implies $P_s k_{s \to s'} = P_{s'} k_{s' \to s}$, where $P$ is the probability of the state denoted by its subscript and $k$ is the transition rate of the processes denoted by its subscript. This expression together with the equilibrium values of the probabilities, $P_s / P_{s'} = e^{-(\Delta G(s) - \Delta G(s'))/RT}$, leads to the following relationship between the probability transition rate constants between two states:

$$k_{s \to s'} = k_{s' \to s} e^{-(\Delta G(s') - \Delta G(s))/RT} .$$

The remarkable property of this expression is that reactions with known rates can be used to infer the rates of more complex reactions from the equilibrium properties. For instance, the association rate of many regulatory molecules to different DNA sites is practically independent of the particular DNA sequence. The dissociation rate, in contrast, strongly depends on the sequence. In this case, knowing one association rate can be used to obtain the dissociation rates for different binding sites through the principle of detailed balance. The inferred rates can then be used to study the dynamics of the system.

The dynamics can be simplified further to any degree of complexity by following a procedure similar to that considered previously for the free energy: it is also possible to perform an expansion for the kinetics of the system, but now in terms of the number of components that can change simultaneously in a transition. We discuss in detail the case in which only one component can change at a given time: either the component gets into or out of the complex. For each component $i$ we can define *on* ($k_{on}^i$) and *off* ($k_{off}^i$) rates for the "association" and "dissociation" rates, respectively, which in principle will depend on the pre-transition and post-transition states of the complex.

The explicit dynamics can be obtained by considering the change in binary variables as reactions

$$s_i \rightarrow (1 - s_i)$$

with rates

$$r_i = (1 - s_i)k_{on}^i(s) + s_i k_{off}^i(s).$$

The reaction changes the variable $s_i$ to 1 when it is 0 and to 0 when it is 1, representing that the element gets into or out of the complex. The mathematical expression of the transition rate reduces to $k_{on}^i$ when the element is outside the complex ($s_i = 0$) and to $k_{off}^i$ when the element is inside the complex ($s_i = 1$). Typically, the *on* rate does not depend as strongly on the state of the complex as the *off* rate. The *on* rate is essentially the rate of transferring the component from solution to the complex. The *off* rate, in contrast, depends exponentially on the free energy. The principle of detailed balance can be used to obtain the *off* rates from the *on* rates:

$$k_{off}^i(s) = k_{on}^i e^{-(\Delta G(s') - \Delta G(s))/RT}.$$

These remarkably compact expressions for the transition rates between different states of the complex can be considered together with other reactions that affect or depend on the state of the complex. In this way, it is possible to integrate the stochastic dynamics of macromolecular assembly into networks of chemical reactions and move the effects of macromolecular assembly up to the properties of cellular processes. The stochastic dynamics of the resulting networks of reactions and transitions can then be obtained with well-established Monte-Carlo algorithms (Bortz et al., 1975; Gillespie, 1976).

## Phage λ cI repressor self-regulation

We illustrate the integration of macromolecular assembly into the dynamics of cellular processes through the kinetics of phage λ cI repressor's self-regulation feedback. The traditional approach to simulate the dynamics of this system would have to consider all the 128 states of the cI-DNA complex as chemical species. Therefore, considering just the kinetics of binding, unbinding, and looping, would involve 128 rate equations, one for each state. In addition to writing down these equations, the traditional approach would need as inputs the values of the 8,128 rates that connect the 128 states with each other. The onerous complexity of the resulting procedure has prevented so far the simulation of the induction switch. There are kinetic studies only for the induction without DNA looping (Arkin et al., 1998). With the approach we have developed, the simulation of the stochastic kinetics of phage λ cI repressor's self-regulation feedback follows straightforwardly.

We have considered the cI-DNA complex together with the different stages of protein production from transcription at the promoter to protein dimerization for active repressors to study the stochastic dynamics of the phage λ repressor self-regulation. The *on* rates of the different transitions are $k_{on}^i = a[N]$ for $i$ = r1, r2, r3, l1, l2, and l3 (describing binding to the DNA operators by cI repressor dimers, cI$_2$) and $k_{on}^i = b$ for $i = L$ (DNA looping), with $a$ and $b$ constants. The *off* rates follow from the detailed balance principle, as delineated previously. In addition, we consider the following chemical reactions (see Figure 5a legend for details): cI mRNA production and degradation; cI repressor production and degradation; cI repressor dimerization and cI$_2$ dissociation; and cI$_2$ nonspecific binding and degradation. The rate of cI mRNA production, $k_t$, is a function of the state of the macromolecular complex, as described previously: $k_t = (k_{act} s_{r2} + k_{bas})(1 - s_{r3})$, where $k_{act}$ and $k_{bas}$ are the activated and basal cI mRNA production rates.

This self-regulatory network incorporates the macromolecular complexity at the promoter region as a module (Fig. 5a). Only a few of the elements of the DNA-protein complex are directly coupled to the cellular dynamics. Specifically, there is an input, the cI dimer concentration ($[N]$), and two main outputs, the occupancy of the r3 and r2 sites, which control the production of cI mRNA ($k_t$). Depending on the free energy of DNA looping, this module has different behaviors (see Fig. 5b-d). For a high free energy of DNA looping (green curves), so that it is very difficult to form the loop, the steady state cI$_2$ concentration (Fig. 5b) is relatively high and noisy in the lysogenic state, when the degradation rate of cI is low (Fig. 5c). In contrast, for free energies of DNA looping close to wild type levels (red curves), $cI_2$ concentration in the lysogenic state is tightly regulated and remains narrowly constrained at low values, exhibiting little fluctuations. Quantitatively, the fluctuations of concentration, usually referred to as noise, are characterized by the variance divided by the mean of the number of molecules: $\eta = \left( \left\langle N^2 \right\rangle - \left\langle N \right\rangle^2 \right) \Big/ \left\langle N \right\rangle$ (Elowitz et al., 2002). In the lysogenic state, DNA looping substantially lowers the strength of the intrinsic noise, from values of $\eta = 7.9$, when there is no DNA looping, to $\eta = 2.7$. Switching from the lysogenic to the lytic states happens as the degradation rate of cI is sharply increased. After the switch, $cI_2$ concentration goes to almost zero in both cases.

The concentration of $cI_2$ also affects another important output of the $cI_2$-DNA assembly module: the occupancy of r1 and r2. This output controls the $P_r$ promoter, which leads to transcription at a given rate when r1 and r2 are free and to no transcription when r1 or r2 are occupied. The activity of a reporter gene controlled by this promoter does not show a marked dependence on the free energy of looping (Fig. 5d). Therefore, DNA looping allows for tight control of $cI_2$ concentration at low levels without substantially affecting the turning on and off of genes at the $P_{rm}$ and $P_r$ promoters. Remarkably, it has recently been found experimentally that turning on transcription of the $P_r$ promoter by increasing the degradation of $cI_2$ is not affected by the presence of DNA looping (Svenningsen et al., 2005). As in the case of the *lac* operon, here DNA looping makes the system function reliably with low numbers of molecules.

## Conclusions: from molecules to networks

The study of cellular processes requires a balance of scope and detail. Whereas single molecular interactions can be modeled in full atomic detail with current technologies (Karplus and Kuriyan, 2005; Saiz and Klein, 2002), approximations in terms of chemical reactions are needed when turning to processes that reach the cellular scale and involve networks of interacting molecular species (Endy and Brent, 2001; Hasty et al., 2002). There are, however, many cellular processes, such as macromolecular assembly, that cannot naturally be described in terms of chemical reactions. One of the major challenges of current biology is therefore to incorporate the molecular details into the description of the dynamics of cellular processes.

We have presented here an approach for bridging the gap between molecular properties and the dynamics of networks of interactions. It can be applied in general to compute the stochastic dynamics of macromolecular assembly networks and their integration into cellular networks. This method is based on a binary description of the potential states of the system and a decomposition of the free energy into a combination of a small subset of elementary contributions of the different components. Such decomposition not only brings forward an extra level of regulation but also provides a starting point to characterize and predict the collective properties of macromolecular complexes, such as looped DNA–protein complexes, in terms of the properties of their constituent elements. The thermodynamic grounds of the method allow for the use of the principle of detailed balance to obtain rate constants, which prevents the appearance of the unrealistic situations noted in existing approaches (Bray and Lay, 1997; Lok and Brent, 2005).

The two examples explored here, the induction switches in the *lac* operon and in phage λ, represent perhaps the most elementary gene regulatory networks of the most basic organisms. And yet, to fully understand the transcriptional regulation in both systems one has to consider thermodynamic quantities that extend beyond standard theory of chemical reactions: macromolecular assembly networks have mechanisms built in that can be used to increase specificity and affinity simultaneously and, at the same time, to control the inherent stochasticity of cellular processes. In particular, exploiting the flexibility of DNA (Cloutier and Widom, 2004; Saiz et al., 2005), protein-DNA complexes can lead to the suppression of cell-to-cell variability, the control of transcriptional noise, and the activation of cooperative interactions on demand (Vilar and Leibler, 2003; Vilar and Saiz, 2005). DNA-protein complexes take full advantage of the conformational properties of DNA by introducing long-range interactions, thus making DNA an active participant in the delivery of the information it encodes.

The mathematical part of the method we have presented has a direct correspondence with a graphical network description, where nodes represent whether or not an element or a property of the component is present, and where the strength of the links between nodes carries information about the effects of these elements on the stability of the complex. Therefore, our approach accounts for the precise stochastic biochemical dynamics, as shown by the prototype systems considered here, while keeping the simplicity of qualitative methods based on network diagrams. This methodology thus offers a solid starting point to move from qualitative to quantitative understanding of protein-protein (Jansen et al., 2003), protein-DNA (Edwards et al., 2002), and other interaction networks (Shen-Orr et al., 2002; Tavazoie et al., 1999) on a genomic scale.

# Figure legends

**Figure 1:** *Representation and thermodynamics of macromolecular assembly.* **(a)** An example of a macromolecular complex in one of eight possible states is used as illustration of the binary description. The complex consists of three molecular positions A, B, and C, described by binary variables $s_A$, $s_B$, and $s_C$, respectively. In this case, A and C are occupied (gray shapes) and B is unoccupied (white shape with dashed contour). Red lines represent pairwise interactions between the components. This description can easily be connected to the thermodynamic properties of different configurations. Here, the free energy $\Delta G^o$ for the configurations and their contributions are expressed in units of kcal/mol. The positional and interaction free energies are assumed to be 15 and -20, respectively. Note that the description refers to a three-molecule complex at a specific location. If, instead, one molecule is used as a reference, its positional free energy should not be counted. The free energy of the disconnected configuration **(b)** is much higher than the free energy of the connected configuration **(c)**. These energetic considerations indicate that the disconnected configuration is extremely less abundant than the connected one. The stability of a compact structure **(d)** is considerably higher than that of a chain-like structure **(c)** because of the additional free energy of interaction.

**Figure 2:** *Interaction networks.* The state and properties of the macromolecular structure can be described by an interaction network. Nodes (big gray circles) in the interaction network represent whether or not a component is present. Small black circles are joined to nodes and represent interactions between the elements they join. Labels associated with black circles indicate the contributions to the free energy arising when all the nodes

they are linked to are occupied. This graphical representation is completely equivalent to the mathematical expression of the free energy in terms of binary variables. For the network shown here, the free energy of a state $s = (s_A, s_B, s_C)$ is given by $\Delta G(s) = p(s_A + s_B + s_C) + e_{AB}s_A s_B + e_{BC}s_B s_C + e_{CA}s_C s_A$, where $p$ is the positional free energy and $e_{AB}$, $e_{BC}$, and $e_{CA}$ are the interaction free energies between the different components. Instantiating all the possible values of the state variable s leads to 8 states, which have a Markov Chain (Norris, 1997) graphical representation in which nodes indicate each specific state $s_A s_B s_C$ of the complex and arrows indicate transitions from one state to another. Only transitions in which a component gets in or out of the complex are displayed here.

**Figure 3:** *Repressor-DNA assembly and gene regulation in the lac operon.* **(a)** The *lac* repressor's binding to DNA can be described by a network of interacting binary elements. Gray circles represent the two DNA-binding domains of the *lac* repressor bound at the DNA sites $O1$ (1) and $O2$ (2) and the lines and black small circles represent interactions. The gray polyhedron represents whether or not DNA is looped (L). $\Delta G(s)$ is the expression of the free energy of the complex in terms of the binary variables. **(b)** The cartoon illustrates the *lac* repressor (green) bound to looped DNA (orange), with the circles and the polyhedron indicating the repressor-operator DNA binding sites and the DNA loop, respectively. The binary variables, $s_1$ and $s_2$, are 1 if the corresponding repressor-operator interaction takes place and are 0 otherwise. Contributions to the free energy (in units of kcal/mol) of the complex include positional ( $p = 15 + 0.6\ln[N]$; considering $\Delta G^o_{pos} \approx 15$ (Finkelstein and Janin, 1989) and $RT = 0.6$ ) and repressor-DNA

interaction ($e_1 = -28.1$ and $e_2 = -26.6$) terms. Here, $[N]$ is the *lac* repressor concentration expressed in moles. For the looped DNA complexes ($s_L = 1$) there are also contributions from the cost of DNA looping ($c_L = 23.35$) and the interaction between sites 1 and 2 mediated by DNA looping ($-p$). This subtraction of a positional free energy in the looped state accounts for the fact that the simultaneously binding of a single repressor to both operator sites should include in the free energy two interaction terms and just one positional term. The free energies used here have been obtained from Ref. (Vilar and Leibler, 2003). To convert from the *in vivo* natural units (molecules/cell) to the more common biochemical ones (concentration) we have used 1 molecule/cell = 1.5 nM. The looping free energy has been computed as described in the caption of Table 1 in Ref. (Vilar and Saiz, 2005) with $\Delta G_{pos}^o = 15$ kcal/mol. **(c)** The free energies $\Delta G(s)$ (in kcal/mol) of the eight different states of the macromolecular network representation of the *lac* operon as a function of the *lac* repressor concentration indicate that the free energy of three states [$s=(0,0,1)$, $(0,1,1)$, and $(1,0,1)$, with $s = (s_1, s_2, s_L)$] is too high and that only the other five states play a relevant role. **(d)** The repression level (R) as a function of the *lac* repressor concentration for one (green circles and dashed lines) and two (red squares and continuous lines) operators shows an excellent agreement with the available experimental data. The values computed with

$$R = [\tau / \tau_{max}]^{-1} = [\frac{1}{Z}\sum_s (1 - s_1)e^{-\Delta G(s)/RT}]^{-1}$$

(lines) are compared with the experimental data (symbols) from Ref. (Oehler et al., 1994) at two repressor concentrations for three different strengths of the main operator $O1$.

**Figure 4:** *cI repressor-DNA complex formation and its effects on the activity of the $P_{rm}$ promoter in phage λ.* **(a)** The network representation of phage λ cI dimers binding to DNA is displayed together with **(b)** a cartoon of the looped configuration with six cI dimers (in green) bound to looped DNA (in orange) and the $P_{rm}$ and $P_r$ promoters. cI repressor dimers can bind to any of the three sites on the right (r1, r2, r3) and on the left (l1, l2, l3) operator, with positional free energy $p = 15 + 0.6 \ln[N]$ kcal/mol and interaction free energies $e_{ri}$ and $e_{li}$, with i=1,2,3 for the right and left operators, respectively. The values used, in kcal/mol, are $e_{r1} = -27.7$, $e_{r2} = -25.7$, $e_{r3} = -25.2$, $e_{l1} = -28.8$, $e_{l2} = -27.1$, and $e_{l3} = -27.4$. In addition, if two cI dimers are bound to two consecutive sites at the right or left operators, there is a cooperative free energy of interaction between dimers, in units of kcal/mol, of $e_{r12} = -3.0$ (r1, r2) and $e_{r23} = -3.0$ (r2, r3) [right operator] and of $e_{l12} = -2.5$ (l1, l2) and $e_{l23} = -2.5$ (l2, l3) [left operator], represented by the lines connecting the neighboring pairs of sites. These two terms have to be compensated in the case that there is one dimer bound to each of the three sites in the left and/or right operators. This is indicated by the line connecting the three sites at each operator with an interaction free energy term of $e_{r123} = 3.0$ kcal/mol (right) and $e_{l123} = 2.5$ kcal/mol (left), and as the diagram indicates this only happens when the three binary variables are $s_{r1} = s_{r2} = s_{r3} = 1$ or $s_{l1} = s_{l2} = s_{l3} = 1$. The additional binary variable $s_L$ indicates whether DNA is looped, which contributes to the free energy with $c_L = 21$ kcal/mol (free energy of DNA looping). DNA looping can mediate interactions between the dimers bound at the left and right operators and, thus, if $s_{r1} = s_{r2} = s_{l1} = s_{l2} = 1$, the cI dimers at both operators can interact and form an octamer, with a contribution to the free energy of the complex of $e_O = -21.5\,\text{kcal/mol}$ and if $s_{r3} = s_{l3} = 1$, the two dimers can interact and form a tetramer with an additional free energy contribution of $e_T = -3.0\,\text{kcal/mol}$. The values of the interaction free energies have been taken from (Dodd et al., 2004) and modified slightly to improve the agreement with the experimental data. The octamer interaction and looping free energies have been chosen so that $c_L + e_O = -0.5\,\text{kcal/mol}$ and their precise value does not affect the results provided that the preceding relationship between them holds. $\Delta G(s)$ is the expression of the free energy of the complex in terms of the seven binary variables. **(c)** The activity of the $P_{rm}$ promoter as a function of the cI dimer concentration is obtained from

$$A = \frac{1}{Z}\sum_s (\tau_{act}s_{r2} + \tau_{bas})(1 - s_{r3})e^{-\Delta G(s)/RT}, \text{ with } \tau_{bas} = 45 \text{ and } \tau_{act} = \tau_{actnl}(1 - s_L) + \tau_{actl}s_L,$$

with $\tau_{actnl} = 420$ for unlooped DNA and $\tau_{actl} = 200$ for looped DNA (Dodd et al., 2004). The cI$_2$ concentration of wild type lysogens, $[N_{lys}] = 2 \times 10^{-7}$ M, is used as reference. The computed activities (full lines) are compared with the experimental data (red symbols) from Ref. (Dodd et al., 2004) for the wild type (squares) system and two other cases: one with a weak l3 binding site with $e'_{l3} = -21\,\text{kcal/mol}$ (triangles) and the other with a large free energy of looping ($c_L = \infty$), equivalent to a system with no left operator (circles).

**Figure 5**: *Phage λ cI self-regulation kinetics.* **(a)** The phage λ cI-DNA assembly network can we viewed as a module within the network that controls the production of cI repressors at the $P_{rm}$ promoter. Binding of cI dimers (cI$_2$) to r3 ($s_{r3} = 1$), represses transcription, whereas binding to r2 when r3 is unoccupied ($s_{r2} = 1$ and $s_{r3} = 0$) activates transcription. The transcription rate is given by $k_t = (k_{actnl}(1 - s_L)s_{r2} + k_{actl}s_L s_{r2} + k_{bas})(1 - s_{r3})$, with $k_{actnl} = 0.05$ s$^{-1}$, $k_{actl} = 0.0225$ s$^{-1}$, and $k_{bas} = 0.005$ s$^{-1}$; cI mRNA is degraded and translated into cI protein at rates $k_{mdeg} = 0.005$ s$^{-1}$ and $k_p = 0.05$ s$^{-1}$ per mRNA, respectively; and cI monomers and dimers are degraded at a rate $k_{deg}$ . The cI dimerization reaction takes place with association and dissociation rate constants $k_a = 7 \times 10^8$ M$^{-1}$ s$^{-1}$ and $k_{dis} = 9.9$ s$^{-1}$, respectively. cI dimers in solution can bind nonspecifically to DNA, with an equilibrium binding constant $K_{ns} = 2.0 \times 10^4$ M$^{-1}$, or bind to one of the free operators with an association rate $k_{on} = a[N]$, with $a = 7 \times 10^8$ M$^{-1}$ s$^{-1}$ . Note that the values used for the association rates are the typical ones of diffusion-limited reactions. The dissociation rates $k_{off}$ depend on the free energy difference between the complexes with (s) and without (s') the cI dimer via the detailed balance principle: $k_{off} = k_{on}e^{-(\Delta G(s') - \Delta G(s))/RT}$ . As cellular volume we have taken 10$^{-15}$ l. The rates of looping and unlooping DNA are $k_{on}^L = 30$ s$^{-1}$ (Vilar and Leibler, 2003) and $k_{off}^L$, respectively, with $k_{off}^L$ obtained also from the detailed balance principle.

**(b)** Time behavior of the number of cI dimers for the case with wild type operators (red) and with no left operator (green) when **(c)** the cI degradation rate $k_{deg}$ (in units of min$^{-1}$) is switched from 0.025 min$^{-1}$ to 0.4 min$^{-1}$ .Time on the horizontal axis is given in hours.

**(d)** Activity of the promoter P$_r$ controlling a reporter gene. In this case, there is transcription at a rate $k_t^{rep} = 0.12(1 - s_{r1})(1 - s_{r2})$ s$^{-1}$ only when both r1 and r2 are free. The reporter mRNA translation, mRNA degradation, and protein degradation rates are $k_p^{rep} = 0.01$ s$^{-1}$, $k_{mdeg}^{rep} = 0.005$ s$^{-1}$, and $k_{deg}^{rep} = 0.005$ s$^{-1}$, respectively.

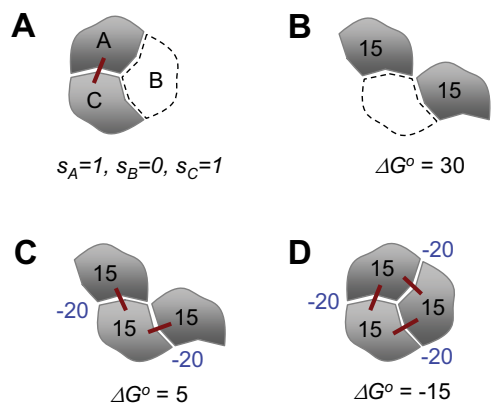

**A**  $s_A=1, s_B=0, s_C=1$

**B**  $\Delta G^o = 30$

**C**  $\Delta G^o = 5$

**D**  $\Delta G^o = -15$

Figure 1

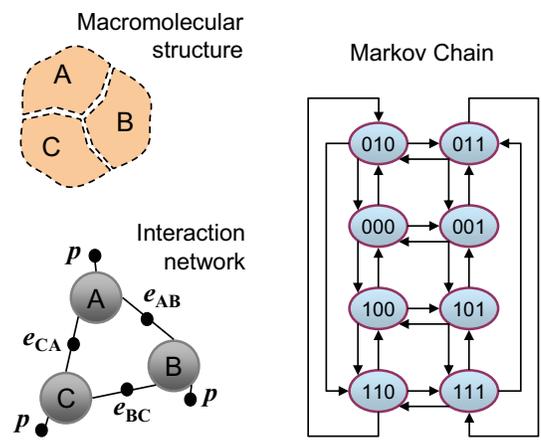

Figure 2

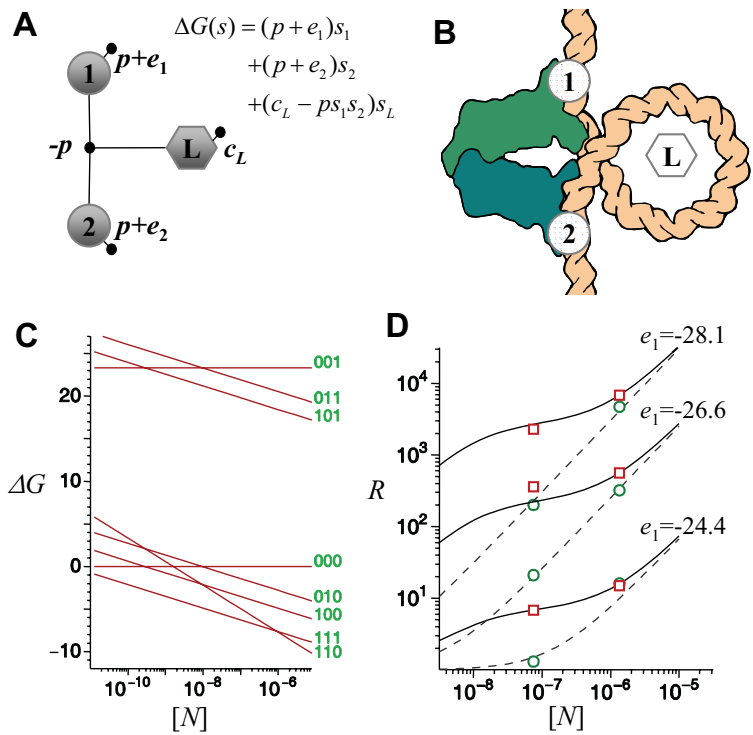

**A**

$$\Delta G(s) = (p + e_1)s_1$$
$$+ (p + e_2)s_2$$
$$+ (c_L - ps_1 s_2)s_L$$

**B**

**C**

$\Delta G$

001
011
101

000
010
100
111
110

$[N]$

**D**

$R$

$e_1 = -28.1$
$e_1 = -26.6$
$e_1 = -24.4$

$[N]$

Figure 3

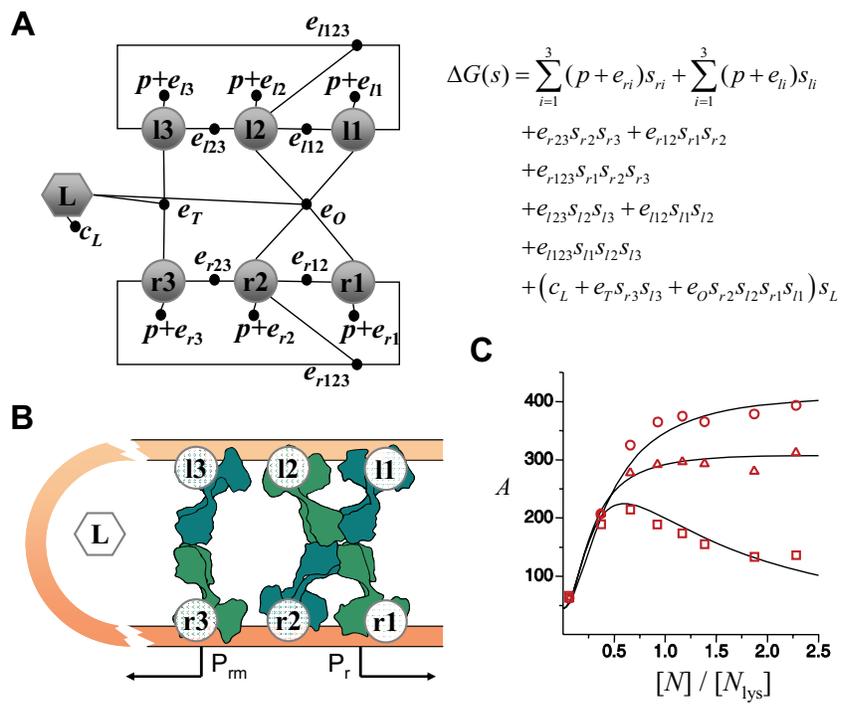

**A**

$$\Delta G(s) = \sum_{i=1}^{3}(p+e_{ri})s_{ri} + \sum_{i=1}^{3}(p+e_{li})s_{li}$$
$$+ e_{r23}s_{r2}s_{r3} + e_{r12}s_{r1}s_{r2}$$
$$+ e_{r123}s_{r1}s_{r2}s_{r3}$$
$$+ e_{l23}s_{l2}s_{l3} + e_{l12}s_{l1}s_{l2}$$
$$+ e_{l123}s_{l1}s_{l2}s_{l3}$$
$$+ (c_L + e_T s_{r3}s_{l3} + e_O s_{r2}s_{l2}s_{r1}s_{l1})s_L$$

**B**

**C**

Figure 4

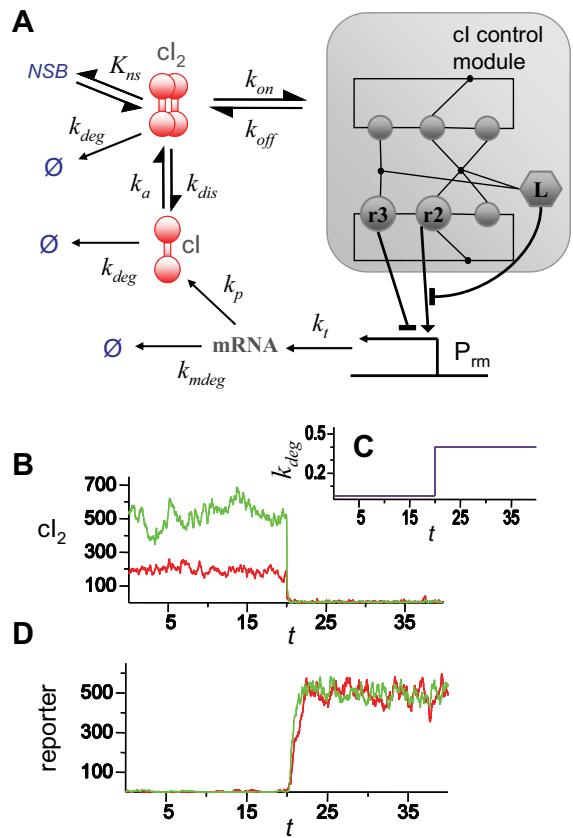

Figure 5